\documentclass[aps,prx,twocolumn,superscriptaddress,noeprint,longbibliography, nofootinbib,floatfix]{revtex4}

\usepackage{graphicx}
\usepackage{bm}
\usepackage{amsmath, amssymb}
\usepackage{booktabs}
\usepackage{mathrsfs}
\usepackage[normalem]{ulem}
\usepackage{pifont}
\usepackage[mathscr]{euscript}

\usepackage{xcolor}

\begin{document}

	\title{Structural magnetic glassiness in spin ice   Dy$_2$Ti$_2$O$_7$ }
	
	\author{Anjana M. Samarakoon}
	\address{Materials Science Division, Argonne National Laboratory, Argonne, IL, USA}
	\address{Shull Wollan Center - A Joint Institute for Neutron 
		Sciences, Oak Ridge National Laboratory, TN 37831, USA}	
	
	\author{Andr\'e Sokolowski}
	\address{Helmholtz-Zentrum Berlin f{\"u}r Materialien und Energie, D-14109 Berlin, Germany}
	
	\author{Bastian Klemke} 
	\address{Helmholtz-Zentrum Berlin f{\"u}r Materialien und Energie, D-14109 Berlin, Germany}
	
	\author{Ralf Feyerherm}
	\address{Helmholtz-Zentrum Berlin f{\"u}r Materialien und Energie, D-14109 Berlin, Germany}
	
	\author{Michael Meissner}
	\address{Helmholtz-Zentrum Berlin f{\"u}r Materialien und Energie, D-14109 Berlin, Germany}
	
	\author{R. A. Borzi}
	\address{Instituto de F\'{\i}sica de L\'{\i}quidos y Sistemas Biol\'ogicos, UNLP-CONICET, La Plata, Argentina}
	
	\author{Feng Ye}
	\address{Neutron Scattering Division, Oak Ridge National Laboratory, Oak Ridge, TN 37831, USA}
	
	\author{Qiang Zhang}
	\address{Neutron Scattering Division, Oak Ridge National Laboratory, Oak Ridge, TN 37831, USA}
	
	\author{Zhiling Dun}
	
	\address{Department of Materials Science and Engineering and Department of Physics and Astronomy, University of Tennessee, Knoxville TN 37996, USA}
	
	\author{Haidong Zhou}
	
	\address{Department of Materials Science and Engineering and Department of Physics and Astronomy, University of Tennessee, Knoxville TN 37996, USA}
	
	\author{T. Egami}
	\address{Shull Wollan Center, Oak Ridge National Laboratory, TN 37831, USA}	
	\address{Materials Science and Technology Division, Oak Ridge National Laboratory, Oak Ridge TN 37831,USA}
	\address{Department of Materials Science and Engineering and Department of Physics and Astronomy, University of Tennessee, Knoxville TN 37996, USA}
	
	\author{Jonathan N. Hall\'en}
	\address{TCM Group, Cavendish Laboratory, University of Cambridge, Cambridge CB3 0HE, UK}
	\address{Max Planck Institute for the Physics of Complex Systems, 01187 Dresden, Germany}
	
	\author{Ludovic Jaubert}
	\address{CNRS, Universit\'e de Bordeaux, LOMA, UMR 5798, 33400 Talence, France}
	
	\author{Claudio Castelnovo}
	\address{TCM Group, Cavendish Laboratory, University of Cambridge, Cambridge CB3 0HE, UK}
	
	\author{Roderich Moessner}
	\address{Max Planck Institute for the Physics of Complex Systems, 01187 Dresden, Germany}
	
	\author{S. A. Grigera}
	\email{sag@iflysib.unlp.edu.ar}
	\thanks{corresponding author}
	\address{Instituto de F\'{\i}sica de L\'{\i}quidos y Sistemas Biol\'ogicos, UNLP-CONICET, La Plata, Argentina}

	\author{D. Alan Tennant}
	\email{dtennant@utk.edu}
	\thanks{corresponding author; present address University of Tennessee Knoxville}
	
	\address{Neutron Scattering Division, Oak Ridge National Laboratory, Oak Ridge, TN 37831, USA}
	\address{Quantum Science Center, Oak Ridge National Laboratory, Oak Ridge, TN 37821, USA}
	\address{Shull Wollan Center, Oak Ridge National Laboratory, TN 37831, USA}
	
	\date{\today}
	
	\begin{abstract}
		The origin and nature of glassy dynamics presents one of the central enigmas of condensed matter physics across a broad range of systems ranging from window glass to spin glasses. 
		The spin ice compound Dy$_2$Ti$_2$O$_7$, which  is perhaps best known as hosting a three-dimensional Coulomb spin liquid  with magnetically charged monopole excitations, also falls out of equilibrium at low temperature.  How and why it does so remains an open question. Based on an analysis of low-temperature diffuse neutron scattering experiments employing different cooling protocols alongside recent magnetic noise studies, combined with extensive numerical modelling, we argue that upon cooling, the spins freeze into what may be termed a `structural magnetic glass', without an a priori need for chemical or structural disorder.  Specifically, our model indicates the presence of frustration on two levels, first producing a near-degenerate constrained manifold inside which phase ordering kinetics is in turn frustrated. A remarkable feature is that monopoles act as sole annealers of the spin network and their pathways and history encode the development of glass dynamics allowing the glass formation to be visualized. Our results suggest that spin ice Dy$_2$Ti$_2$O$_7$ provides one prototype of magnetic glass formation specifically, and a setting for the study of kinetically constrained systems more generally.   
		
	\end{abstract}
	
	\maketitle
	
	
	\section{Main Text}
	
	Magnetism provides a remarkable playground for understanding complex phenomena. A wide variety of materials with spin networks that closely 
	match theoretical models as well as the intuitive nature of spin degrees of freedom account for their wide-ranging importance. Spin glasses are a celebrated instance of this as examples of states where equilibrium breaks down. More recently, fractional quasiparticles and spin liquids have come into focus as scenarios where exotic phases emerge from highly frustrated landscapes. The observation of anomalous noise and development of memory effects in the highly frustrated spin-ice compound Dy$_2$Ti$_2$O$_7$ \citep{samarakoon_2021} points to a close connection between these. Given that an essential component of conventional spin glasses, namely randomness, is believed to be absent the prospect of a spin realization of a structural glass instead is a tantalizing prospect; especially given the simplifications a spin model could bring.     
	
	Spin liquids~\cite{balents2010spin,knolle2019field}, unlike paramagnets, are not simply thermally disordered. Rather, they are characterized by a huge density of low-energy  spin states. These can incorporate intricate correlations which emerge from local constraints imposed by competing interactions. At low temperature they can thus be fundamentally different from conventional magnets, as evidenced by their exotic excitations  – such as point-like monopoles or extended Dirac strings~\cite{Castelnovo2008}  – whose collective behavior we are only beginning to understand~\cite{Castelnovo2008,Morris2009,Fennell2009,Castelnovo2012,ross2011quantum,gingras2014quantum}. In particular, their dynamics is subject to the vagaries of the interplay of the unusual nature of the degrees of freedom with the subtle effects of other interaction terms that are inevitably present in real materials and give rise to an intricate energy landscape.  
	
	\begin{figure*}
		\centering
		\includegraphics[width=1.7\columnwidth]{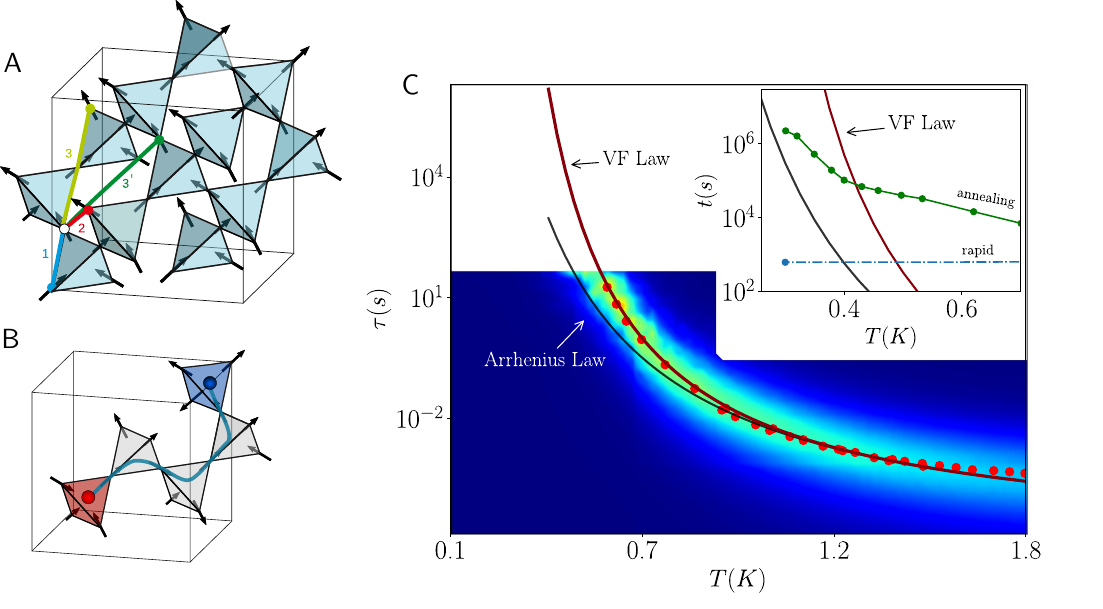}
		\caption{Magnetic structure, monopole hopping and relaxation time of DTO. (A) The spins in DTO are located on a pyrochlore lattice. The magnetic interaction pathways are shown in color, dipolar coupling (not shown) is also important. These constrain the spin configurations to follow two-in two-out ice rules. Breaking the ice rules results in the creation of a monopole anti-monopole pair which can separate, creating a Dirac string as shown in (B). Monopoles are constrained by the other spins (Dirac strings) to travel over a restricted manifold. (C) Relaxation time ($\tau$) for DTO extracted from noise measurements follows a Vogel-Fulcher-Tammann (VF) behavior below $1.5$~K, rather than an Arrhenius law.  The color map is of $\chi^{\prime\prime} (\omega)$ whose maximum gives a measure of $\tau$. The inset shows the cooling protocol for one cycle of the neutron experiments: the symbols correspond to the waiting times as a function of temperature for the rapid (light blue) and slow cooling (green) protocols of the neutron experiments. These are compared with the VF law obtained from the noise and with the Arrhenius law fit to the data. 
		}
		\label{fig1}
	\end{figure*}
	
	Dy$_2$Ti$_2$O$_7$ (DTO) is a spin ice material that belongs to the family of pyrochlore oxides of type A$_2^{3+}$B$_2^{4+}$O$_7^{2-}$ containing a wide range of highly frustrated magnets, many of which fail to order at low temperatures~\cite{Castelnovo2008,Morris2009,Fennell2009,Henley2010,gardner2010magnetic,Castelnovo2012,rehn2016maxwell}. The magnetic lattice is shown in Fig.~\ref{fig1}A. Ising spins are located on the corners of tetrahedra and are restricted to pointing in- and outward. Recent experimental results demonstrated that a comprehensive fit to several thermodynamic properties~\cite{yavors2008dy,borzi2016intermediate,henelius2016refrustration,Samarakoon_2020} requires a Hamiltonian with exchange interactions to third nearest neighbors (as indicated in the figure), along with longer-ranged dipolar interactions. The resulting  constrained states mentioned above are described by the ice rules, which stipulate two spins pointing in, and two out, at each tetrahedron~\cite{harris1997geometrical,Bramwell2001}. Mapping spin vectors to an emergent magnetic field B gives a divergence free condition, $\nabla \cdot B =0$, reminiscent of magnetic fields in conventional magnetostatics. Indeed, if the ice rules are broken, say by a thermal single spin flip excitation, then two topological defects (a pair of oppositely charged magnetic monopoles) form; and as they separate, they turn spins over along the way forming a Dirac string [see Fig.~\ref{fig1}B]. Pure spin ice is in a Coulomb phase where the monopoles are deconfined, interacting through an emergent $1/r$ potential, as expected for charges in ``free space'' with the spin correlations having the functional form of free dipoles~\cite{Castelnovo2008}. A central scientific question  is how perturbing pure spin ice and reducing thermal disorder generates complex forms of matter such as glasses. 
	
	In this work, we investigate the regime in which memory effects appear in Dy$_2$Ti$_2$O$_7$, at temperatures {\it much higher} than any predicted underlying ordering temperature~\cite{samarakoon_2021}. We study this history-dependence in a detailed neutron scattering study incorporating different cooling protocols, connecting these results with extant magnetic susceptibility and noise measurements. Our numerical modelling indicates that this `supercooled'-liquid-like behaviour is intimately connected to the passage of monopoles in an energy landscape that is {\it not quite} flat: interactions that at low enough temperatures induce order affect the potential landscape explored by monopoles as they hop from site to site, slowing down their dynamics at the same time as their density plummets. We model the resulting monopole pathways numerically, contrasting the regime where the spin network ultimately become glassy with the behaviour of a robustly ordered magnet.

	\section{Results}
	
	\subsection{Spin interactions and glassiness}
	
	As DTO is cooled, the spin dynamics undergoes drastic slowing down, with the appearance below $T_\mathrm{irr}\approx 600$~mK of history-dependence -- typically evidenced by a divergence between FC and ZFC magnetisation measurements -- and a host of accompanying phenomena such as thermal runaway~\cite{melko2001long,snyder2001spin,snyder2004low,Jaubert2009,slobinsky2010unconventional,Jaubert2011,matsuhira2011,yaraskavitch2012spin,Paulsen2014,jackson2014dynamic,Guruciaga_2020}  The temperature dependence of a characteristic relaxation time, $\tau(T)$, can thus be extracted, with Fig.~\ref{fig1}C showing the values from the most recent magnetic noise experiments \cite{samarakoon_2021}, which are in good agreement with previously determined values (e.g. from  \cite{snyder2001spin,matsuhira2011spin,pomaranski2013absence}). Fitting a functional form is complicated by the temperature dependence of various physical processes involved in the dynamics. 
	An Arrhenius law (black curve) departs significantly from the data below approximately $800$~mK. 
	Instead, $\tau(T)$ can be well-fitted by a Vogel-Fulcher-Tammann (VF) form, $\tau(T)=\tau_0\exp(A/(T-T_{\rm VF}))$, common for glasses (see e.g. \cite{shtrikman,garca1989theoretical}), with $\tau_0 = 7.0(3) \times  10^{-6} s$,  $T_{\rm VF}= 0.18(2) K$ and $A = 6.0(9) K$ (red curve), across all of the accessible temperature range~\footnote{See Sup. mat. for a similar analysis of Monte Carlo simulation data.}. The VF law a priori reflects cooperative effects rather than single-ion physics. $T_{\rm VF}$ is the Vogel divergence temperature, where the system reaches an infinite relaxation time. The figure also shows a color map of $\chi^{\prime\prime}$ obtained from noise measurements~\cite{samarakoon_2021} by means of the fluctuation dissipation theorem, whose maximum also gives a measure of $\tau$. Measurements at temperatures and on timescales below and to the left of $\tau(T)$, respectively, probe the out-of-equilibrium behavior. Around $600$~mK, the relaxation time is close to $200$~s, and it rises swiftly below that temperature, exceeding $10^5 s$ below $300$~mK. 
	
	\begin{figure*}
		\centering
		\includegraphics[width=1.7\columnwidth]{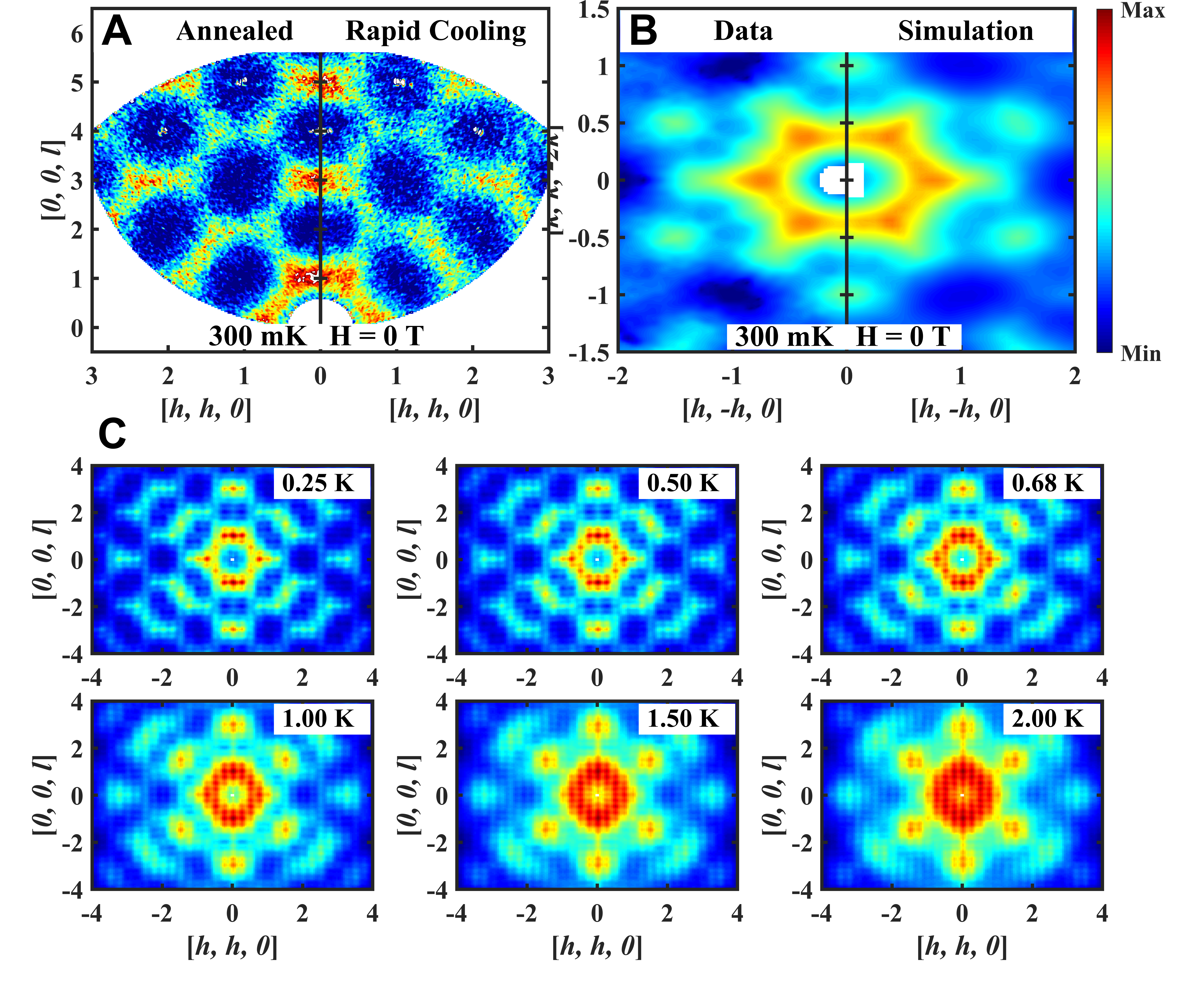}
		\caption{Glass behavior in DTO.  (A) Fast (1 hour) and slow (`annealed' over 4 weeks) cooling data, taken along trajectories indicated on Fig. 1 C (inset), show little discernible difference (the measurements correspond to points A2 and R2 in the sample temperature history given in the Sup. Mat.). (B) Modeled and measured (rapid cooled) diffuse scattering agree at low temperature. 
			See Sup. Mat. for details on the neutron scattering measurements and quantitative line-cut comparisons.
			(C) Simulations of the diffuse scattering  show Coulombic correlations associated with the Coulomb spin liquid at 1 to 2 K. At lower temperatures the build-up in correlations is reflected in a sharpening of the correlations. Below $600$~mK the correlations remain largely unchanged, as expected for a glass. The simulations correspond to a Hamiltonian with dipolar interactions and up to third nearest neighbour exchange interactions with 
			$D = 1.3224$~K, 
			$J_1=3.41$~K, $J_2=0$, $J_3=-0.004$~K and $J_3'=0.044$~K (see Sup. Mat. for details).}
		\label{fig2}
	\end{figure*}

	A previous machine-learning study by a subset of the present authors identified an effective spin-ice Hamiltonian that, in the manner of Refs.~\onlinecite{yavors2008dy,henelius2016refrustration,borzi2016intermediate}, consists of a dipolar interaction combined with exchange terms up to third nearest neighbours (which are of two different kinds in the pyrochlore lattice). A prolate spheroid in parameter space was determined inside which several experimental observations, such as neutron scattering, specific heat and magnetisation, were well described by the model. Notably, this model, when simulated using a single spin flip algorithm (SSF), successfully captures dynamical features of the system, such as the FC-ZFC irreversibility at $600$~mK~\cite{snyder2004low,bovo2018special}, and -- if the standard correspondence between MC steps and real time is used -- it can also reproduce the  departure from Arrhenius behaviour in the magnetic relaxation time scale, as shown in the Sup. Mat. 
	
	To probe the nature of this freezing we measure diffuse neutron scattering spectra under different cooling protocols, ``annealing'' with a timescale of several days, as well as rapid cooling over an hour.  The full temperature history of two independent cooling  cycles used for these measurements are shown in Fig S1 of the Sup. Mat..  Both cycles give identical results, so we focus our discussion on the longer one (cycle 2), with an annealing time close to a month (data for cycle 1 is shown in Sup. Mat.).  The inset of Fig.~\ref{fig1}C shows in logarithmic scale the waiting times at each temperature for each cooling protocol in this cycle, compared with the characteristic times from the VF and Arrhenius laws, as the temperature is lowered towards $300$~mK. According to the VF law, even the longest waiting time, which slightly exceeds those of previous long-cooldown experiments (\cite{pomaranski2013absence,giblin2018}), are out of equilibrium below 450 mK.  The data from the neutron scattering experiments supports this statement; as seen in (Fig.~\ref{fig2}A) rapid and annealed cool-downs give indistinguishable scattering patterns (see Sup. Mat. for a comparison and subtraction of raw data for cycle 1); this is also in agreement with the results seen in Ref.~\onlinecite{giblin2018}. 
	
\begin{figure*}
		\centering
		\includegraphics[width=2.0\columnwidth]{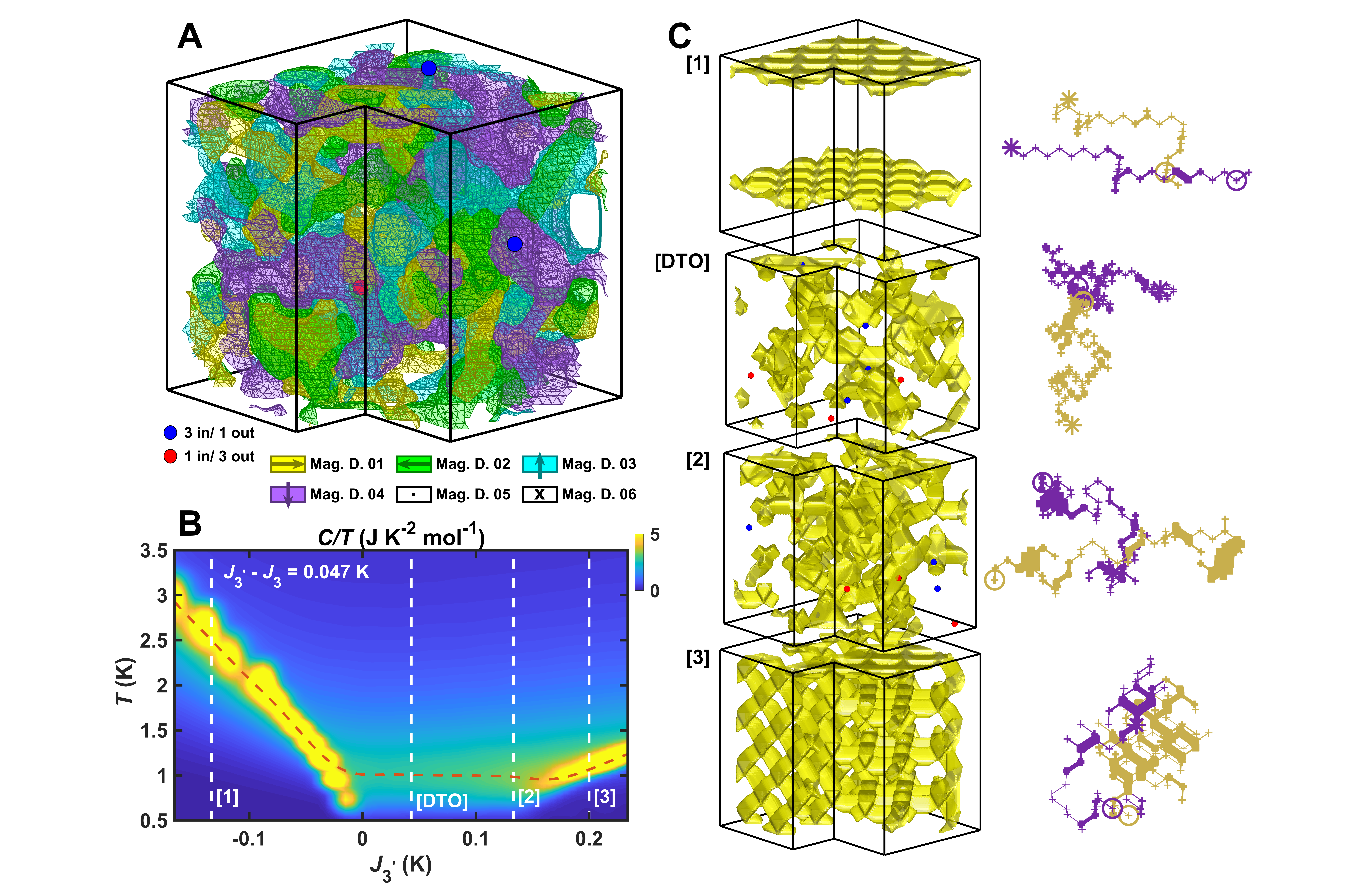}
		\caption{Spin configurations and monopole trajectories as a function of 3rd neighbor interaction strength. (A) Connected tetrahedra with the same net magnetic moment define ferromagnetic (FM) regions, depicted by the smooth surface wrapping around them. Ice rules allow six FM orientations. The configuration here is modeled at $T= 300{\rm ~mK} < T_{\rm irr}$ using parameters from Ref.~\onlinecite{Samarakoon_2020}, see text. For clarity, only four FM orientations are shown. Monopoles are trapped in this structure, as can be seen in the animation of the time dependence presented in SI (Movie S1). (B) Calculated heat capacity over temperature ($C_v/T$) along the line $J_3-J_3\prime = 0.047$~K plotted as a function of $J_3\prime$ and $T$. Order is seen at either limit, interrupted by glass formation in a broad intermediate region. The glass formation occurs at a near constant temperature which disrupts order at its characteristic temperature. C) The morphology of the FM regions changes with 3rd NN interactions. A single type of FM region is shown for the parameters indicated by vertical dashed lines in (B). This shows the influence of different ordering tendencies. To the right of each sub-figure, a pair of corresponding typical monopole pathways are shown.}
		\label{fig3}
	\end{figure*}

	We find that the effective Hamiltonian gives an accurate quantitative description of the experimental NS data, both after rapid coooling and annealing (Fig.~\ref{fig2}B). In accordance with the experiments, simulations of the diffuse scattering show Coulombic correlations associated with the emergent $U(1)$ gauge field between $2$~K and $1$~K (Fig.~\ref{fig2}C). At lower temperatures, a further build-up in correlations is reflected in a sharpening of the features. Below $600$~mK the correlations remain largely unchanged as one would expect for a glass, with no appreciable features indicating the development of long-range order,  as opposed to numerical simulations using non-local updates where Bragg peaks develop at very low temperatures~\cite{melko2001long,borzi2016intermediate,yavors2008dy,henelius2016refrustration,bovo2018special}.
	A calculated spin configuration below $T_{\rm irr}$ is shown in Fig.~\ref{fig3}A, whose structure factor matches the experimental NS data accurately (Fig.~\ref{fig2}B). To visualise the spatial magnetic correlations in the system, we use colored smooth surface wrappings around regions of connected tetrahedra with the same net magnetic moment. Ice rules allow six possible such  orientations. For clarity, only four types are shown.  
	
	While computationally intensive, effective Hamiltonians open the possibility of investigating possible mechanisms by which the cooperative contribution to spin freezing might occur.

	\subsection{Monopole motion and domain formation}

\begin{figure}
		\centering
		\includegraphics[width=\columnwidth]{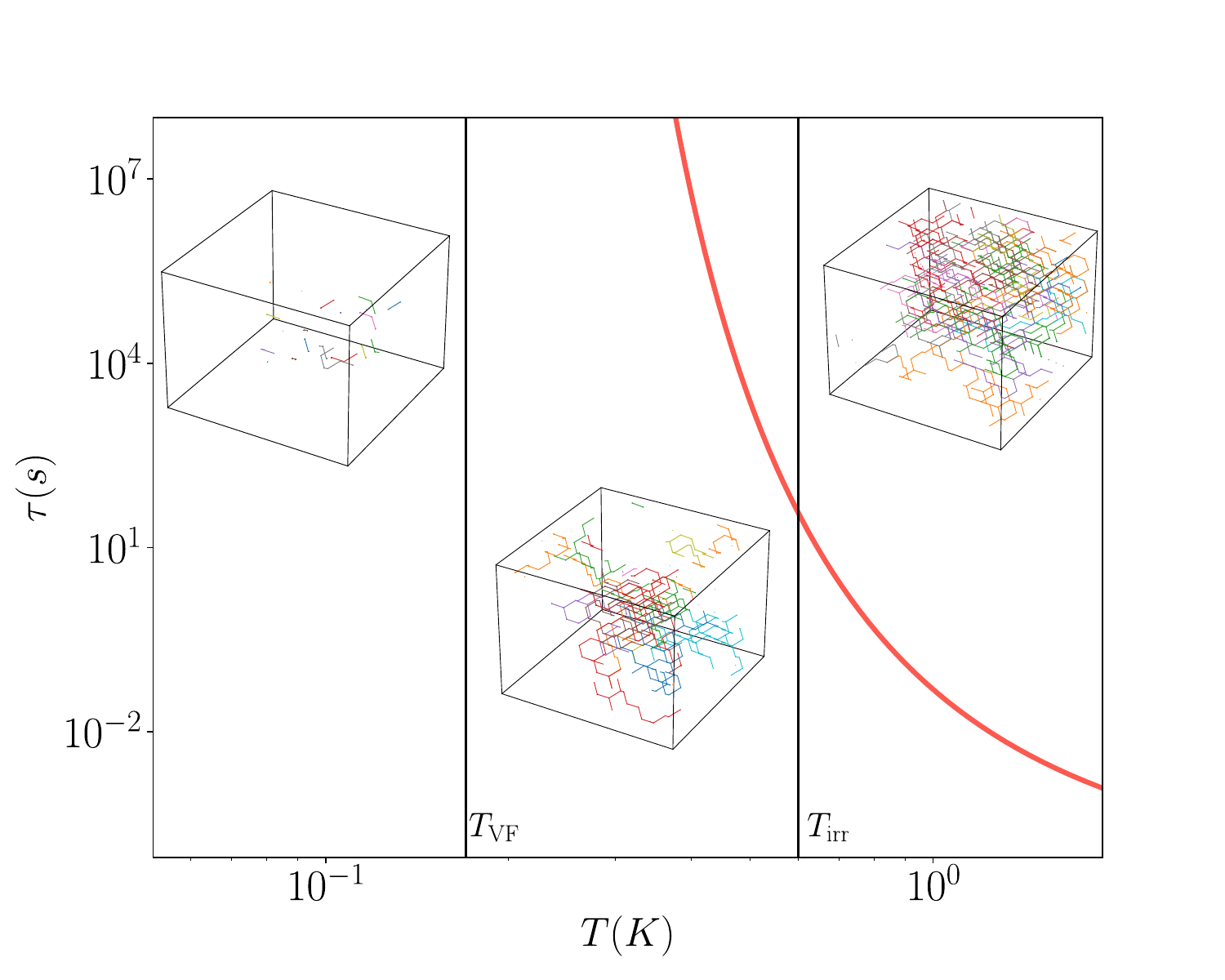}
		\caption{Monopole pathways at different temperatures.  For clarity of comparison, the same (artificially) fixed density of monopoles is shown at each temperature; in reality, the number of monopoles at lowest temperature is vanishingly small. The graph shows characteristic monopole paths taken over a finite MC time in three regions.  Above the irreversibility line, $T > T_{\rm irr}$, the monopole paths are extended and mutually connected and (eventually) span the whole sample. For $T_{\rm VF} < T < T_{\rm irr}$, the monopole pathways take the form of sparse networks; 
			as temperature is lowered their size and dimensionality is increasingly reduced and they become disconnected. For $T < T_{\rm VF}$ the monopoles are essentially frozen and the pathways become zero dimensional.  The red line is the VF law that best fits the relaxation time data (shown in Fig. 1).
		}
		\label{fig4}
	\end{figure}

	The simulations show spin dynamics in three distinct regions (illustrated in Fig.~\ref{fig4}).
	In the liquid, $T>T_{\rm irr}$, monopole pathways overlap and are three dimensional. Short-range ferromagnetic clusters undergo bursts of rapid change as they are traversed by monopoles; over time the entire simulated volume reaches equilibrium (see Movie S1). In the intermediate region ($ T_{\rm VF} < T < T_{\rm irr}$) the monopoles travel on sparse networks. The monopoles are then confined in intertwined FM regions, with  the simulation timescale shorter than that on which large scale spin rearrangements happen. Here monopole motion causes only local fluctuations of the clusters but not volume equilibration  (Movie S2). At the lowest temperatures, ($T < T_{\rm VF}$), the pathways are short and spatially confined, i.e., effectively zero dimensional. The monopoles have been reduced to short range hops and the structure has hardened (Movies S3 and S4). 
	
	Supporting quantitative analysis including monopole lifetimes, displacement, fractional visitation of sites, jumping rates, as well as quantification of looping of the pathways (using the first Betti number $N_{\rm Betti}$~\cite{hatcher2002,Betti}) is provided in the Sup. Mat..

	Our choice to represent magnetic correlations in terms of FM regions preserves the full microscopic information of the ice states, but does not reflect the low-temperature ordered state of the model, which is of the Melko-Gingras type (see Refs.~\onlinecite{melko2001long,borzi2016intermediate,henelius2016refrustration}). We note, however, that monopole motion is constrained to follow Dirac strings, i.e., it occurs preferentially along/opposite to (depending on the sign of the charge) the direction of the ferromagnetic moments, so that the time-dependence of the FM domains provides a natural window into the dynamical processes in these systems. 
	
	Interactions modify the short-range structures and therefore the character and effective geometry of the pathways. Notably, the third nearest-neighbor interaction plays a significant role in determining the low temperature phase of this material.  Fig.~\ref{fig3}B shows the simulated heat capacity over temperature $C_v/T$ as a function of temperature along a straight line in the $J_3 - J_3\prime$ plane defined by ($J_3\prime - J_3 =0.047$~K), with $J_3\prime$ ranging from $-0.15$~K to $0.22$~K. The order that exists at both ends of the diagram, marked by a peak in $C_v/T$ that decreases in temperature towards the centre, is interrupted in a broad region between $-0.1$~K and $0.3$~K, where the system enters a glassy state at a roughly constant temperature. Our model for DTO lies at the center of this broad glassy region; this is a robust, rather than fine-tuned situation, that gives some latitude in the choice of parameters. 
	The ferromagnetic clustering and, consequently, the character of the monopole pathways are also strongly dependent on $J_3$ and $J_3\prime$.  
	Fig.~\ref{fig3}C shows the distribution  of one region (coloured yellow), alongside the monopole pathways, for different strengths of the third neighbour couplings, marked as dotted lines in the phase diagram of Fig.~\ref{fig3}B. For negative $J_3$ and $J_3\prime$, (panel [1] in Fig.~\ref{fig3}C), the low  temperature phase is ordered, and the domains form parallel planes along two principal axes; as $J_3\prime$ increases towards the value for DTO, (panel [DTO]), the system exhibits a broader  disordered low temperature phase, reflected in the regions becoming rugged and distributed across the whole sample. When $J_3\prime$ is further increased, the regions become progressively ordered, (panel [2] in~\ref{fig3}C), and eventually, when the system orders at low temperature, they form a uniform criss-cross pattern, (panel [3] in~\ref{fig3}C). The right column shows the pathways for two monopoles in each case.  
	
	The frustrated DTO thus sits, as expected, in a region with a complex balance between dipolar, near and further neighbor interactions; here, small changes in temperature significantly alter the morphology of the monopole pathways. This results in different dynamical behavior which is seen in the simulations in a non-Lorentzian noise spectral density with a temperature dependent anomalous exponent. The simulated noise spectrum, with a single spin-flip rate, matches qualitatively but not quantitatively that of the experiments. 
	As discussed in a recent work~\cite{samarakoon_2021}, a detailed modelling of the dynamics requires consideration of both cooperative behaviour and the local physics of the spin-flipping process. While this is necessary to accurately describe the noise spectrum, they do not alter the morphology of monopole pathways. 
	
	\subsection{Physical origin of glass formation}
	
	Our results suggest an interpretation where glass forms from a combination of rarefaction of monopoles and the build-up of short-range correlations in turn restricting their motion. These correlations appear to be brought about by precursors to ordering that would eventually occur in thermodynamic equilibrium at lower temperatures -- thus producing a remarkable instance of frustrated phase ordering kinetics preceding the ordering temperature.
	
	With cooling, large scale annealing events, where many monopoles exchange position in bursts and the effective potential landscape changes akin to an avalanche, are unlikely to be sampled in any experimental time scale. Measurements instead reflect the common short-ranged excursions of monopoles which result in fluctuations around, rather than changes of, a given ice configuration. As temperature decreases, even these motions are increasingly blocked and below $T_{\rm VF}$ annealing is effectively prevented as energy barriers are too large for monopoles to move between energy minima on experimentally accessible time scales.   
	
	Spin glasses are usually found in materials with randomness as well as frustration. Relaxor ferroelectrics are created as a result of strong ionic site disorder~\cite{cross1987relaxor}. DTO on the other hand has no evident intrinsic randomness or disorder. Instead, it has much closer connection to structural glass behavior where networks of bonds, defects, and coordination are important. Some liquids, particularly those with high degrees of frustration, remain liquid down to low temperatures without crystallization, and become glassy due to kinetic slowdown~\cite{debenedetti2001supercooled}. In DTO similar behavior appears to occur because the primary couplings are large compared to the glass formation temperature, and the strong frustration stabilizes the liquid state down to low temperatures. The small perturbation of distant neighbor couplings then freeze the fragile domain structure. The behavior of DTO below 
	800 mK resembles the development of dynamic heterogeneity in supercooled liquids~\cite{ediger2000spatially,glotzer2000time}. 
	Previous measurement of a VF law in spin-ice was associated with a supercooled spin liquid~\cite{kassner2015supercooled}. The phase diagram in Fig.~\ref{fig3}B does not indicate an avoided long-range ordering around $600$~mK, and the equilibrium ordering temperature into a Melko-Gingras type crystal, found using loop algorithms in our model Hamiltonian, lies close to $200$~mK. This temperature, unlike the case of supercooled liquids, coincides within experimental error with $T_{\rm VF}$. This is a consequence of the freezing in the mobility of defects that occurs at crystallisation, which implies a divergence of the characteristic time. 
	
	\section{Discussion and conclusions }
	
	The nature of the low-temperature state of spin-ice materials has been a matter of continued investigation. As $T$ is lowered into the spin ice phase, characteristic Coulomb correlations appear that can be described by a dipolar spin-ice model. Crucially, however, the behavior of the `pure' version of this ice model, and its varieties, differ greatly in this regime. What is special about the dipolar spin ice model is that its constrained manifold of the exponentially many degenerate states obeying the ice rules is very `flat': in a slightly simplified model, such as the dumbbell model~\cite{Castelnovo2008}, all these states are degenerate. The motion of the monopoles, which connects these states, thus takes place without encountering significant energy barriers, and only slows down as the size of the monopole population is thermally suppressed, following an Arrhenius law~\cite{lubchenko2007theory}. Corrections to the  point charge approximation of the monopoles, and further neighbor exchange interactions, however, select a subset of ground states which is generically small and involves spontaneous (at least partial) ordering  at even lower temperatures. It is a much investigated and long-standing issue in the physics of spin ice that such ordering has never been observed experimentally. Our observations advocate the view that these subdominant interactions generate obstacles to the monopole motion by endowing the previously flat energy landscape with additional structure: this is the frustrated kinetics in the constrained manifold arising from the ice rules. It is in this sense that we refer to spin ice as a structural magnetic glass: like a structural glass, its slow, non-Arrhenius, dynamics can be the result not of disorder or boundary effects~\cite{revell2013evidence}, but of a non-trivial energy landscape arising as a cooperative effect. 
	
	The high tunability of magnetic materials like DTO with pressure and chemical substitution, their simple and well-defined interactions, and lattice nature make them highly attractive as a route to understanding the complex physics behind structural glasses.  In these materials, the remarkable physics of partially disordered phases of matter and how they emerge from competition is well suited to machine learning. This can be applied to spin simulations directly and may provide surprising new insights.  
	Highly degenerate manifolds are increasingly being studied in topological and quantum materials and the findings here should be highly relevant. Spin jammed states~\cite{chandra1993anisotropic,samarakoon2016aging,bilitewski2017jammed,bilitewski2018disordered,rau2016spin}
	have been proposed in frustrated magnets, and disordered states appear to compete with potential quantum spin liquids. In addition, narrow band electronic systems are being explored that will also have large degenerate manifolds if tuned to the Fermi energy~\cite{kang2020dirac}. 
	These considerations are not limited to real materials but also extend to artificial magnetic arrays~\cite{Schiffer2021}, where similar instances of slow dynamics and glassiness have been encountered in recent years~\cite{Skjaervo2020}.

	In conclusion, the dipolar and small further neighbor interactions in spin ice Dy$_2$Ti$_2$O$_7$ appear to be responsible for the appearance of glassy dynamics. On cooling, short-range correlations emerge which introduce restrictions on monopole pathways. This eventually breaks the ergodicity of the system on measurement time scales at a temperature matching the irreversibility temperature, $T_{\rm irr} \approx 600$~mK, and a glassy state exhibiting only the short-range correlations is formed. Dynamics gradually becomes yet more sluggish as temperature drops further and our fit of the relaxation time to a Vogel-Fulcher law leads to a temperature scale $T_{\rm VF} \approx 200$~mK for its divergence which is close to the numerically obtained thermodynamic transition temperature into a fully ordered state. The ordering tendencies and geometry of the pathways are shown to be controllable by the strength of further neighbor interactions. 
	
	The simplicity of the model suggests that spin ice presents new opportunities for microscopic understanding of glass formation and a new platform to study glassy behavior.  
	
	
	\section*{Acknowledgements}
	DAT and AMS would like to thank Cristian Batista and Erica Carlson for useful discussions. Funding: ZLD and HDZ thank the NSF for support with grant number DMR-2003117. A portion of this research used resources at Spallation Neutron Source and was supported by DOE BES User Facilities. This material is based upon work supported by the U.S. Department of Energy, Office of Science, National Quantum Information Science Research Centers, Quantum Science Center. Support for QZ was provided by US DOE under EPSCoR Grant No. DESC0012432 with additional support from the Louisiana board of regent. The computer modeling used resources of the Oak Ridge Leadership Computing Facility, which is supported by the Office of Science of the U.S. Department of Energy under contract no. DE-AC05-00OR22725.  SAG thanks Agencia Nacional de Promoci\'on Cient\'{\i}fica y Tecnol\'ogica through PICT 2017-2347. This work was partly supported by the Deutsche Forschungsgemeinschaft under grants SFB 1143 (project-id 247310070) and the cluster of excellence ct.qmat (EXC 2147, project-id 390858490), by the Helmholtz Virtual Institute ``New States of Matter and their Excitations'', and by the Engineering and Physical Sciences Research Council (EPSRC) Grants No. EP/K028960/1, No. EP/P034616/1, and No. EP/T028580/1 (CC). Part of this work was carried out within the framework of a Max-Planck independent research group on strongly correlated systems. TE was supported by the U.S. Department of Energy, Office of Science, Basic Energy Sciences, Materials Sciences and Engineering Division.

	\section*{Author contributions}
	
	Conception of measurement (DAT SAG) and wider project (DAT SAG CC RM). Neutron scattering measurements (SAG AS DAT BK RF MM QZ ZD HZ). Modelling, analysis and interpretation (AS SAG DAT TE RAB JH LJ CC RM).
	Paper writing (RM CC SAG DAT with input from all coauthors). 
	
	Competing interests: The Authors declare no competing interests.
	
	Data and materials availability: The data and material that support the findings of this study are available from the corresponding author upon reasonable request.
	
	\bibliography{references}

%

\newpage
\quad

\newpage

\setcounter{section}{0}
\setcounter{figure}{0}

\renewcommand{\thesection}{S\Roman{section}}
\renewcommand{\thefigure}{S\arabic{figure}}

\section*{Supplementary Material}
	\section{Neutron Scattering experiments under different cooling protocols}
	
	\begin{figure}[h]
		\centering
		\includegraphics[width=\columnwidth]{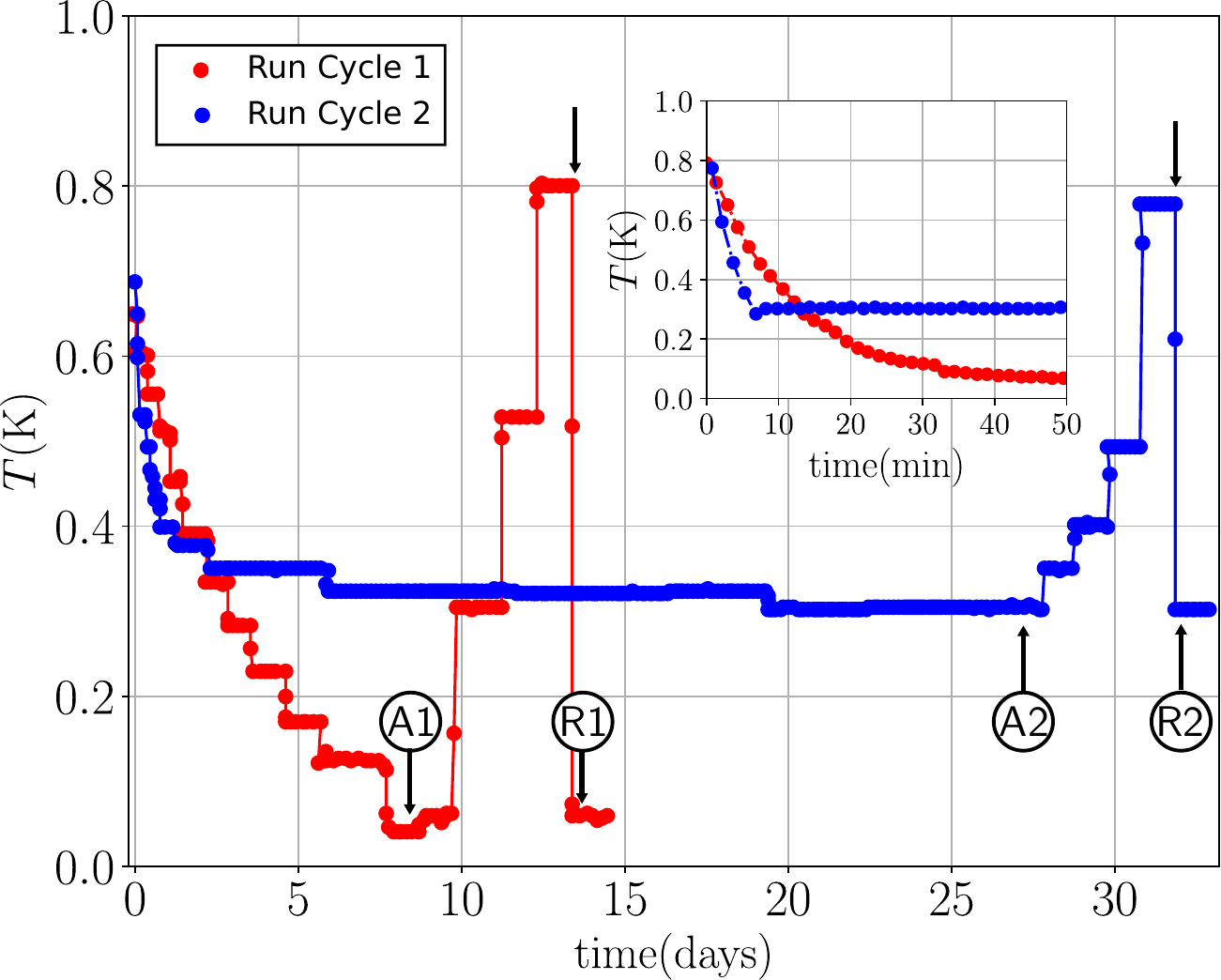}
		\caption{Cooldown procedure of the Berlin neutron scattering measurements: The figure shows the temperature recorded by a thermometer attached to the base of the sample as a function of time (in days) for the two experiments performed at HZB Berlin (red and blue). Cycle 2 (in blue) corresponds to the data shown in the main article (Figs.~1(c) and 2(a)) and cycle 1 to the data shown in Fig.~S2). The labels A1 and A2 correspond to the times when annealed measurements were made, while R1 and R2 indicate the times corresponding to rapid coolings. The inset shows a detail of the temperature as a function of time for these two rapid coolings (the beginning of which is indicated by black arrows in the main panel) with higher temperature sampling.
		}
		\label{S1}
	\end{figure}
	
	Isotopically enriched single crystal samples of Dy$_2$Ti$_2$O$_7$ with $95-98$\% Dy-162 were used to perform diffuse neutron scattering experiments (see Methods: Experimental detail in Ref.~\onlinecite{Samarakoon_2020} for more information). A single crystal of mass $\approx 200$~mg was aligned in the (111) plane for the neutron investigation at the single crystal diffuse scattering spectrometer CORELLI at the Spallation Neutron Source, Oak Ridge National Laboratory. The crystal was prepared as a sphere to minimize absorption corrections and demagnetization corrections. CORELLI is a time-of-flight instrument where the elastic contribution is separated by a pseudo-statistical chopper~\cite{ye2018implementation}. The crystal was rotated through $180$ degrees with the step of $5$ degrees horizontally with the vertical angular coverage of $\pm 8$ degree (limited by the magnet vertical opening) for survey on the elastic and diffuse peaks in reciprocal space. The dilution refrigerator insert and cryomagnet were used to enable the measurements down $100$~mK and fields up to $1.4$~Tesla. The data were reduced using Mantid~\cite{arnold2014mantid} and Python script available at CORELLI. Background runs at $1.4$~Tesla were made to remove all diffuse signals. The extra scattering at Bragg peak positions due to the polarized spin contribution was accounted for using zero-field intensities. Fig.~2(b)(left) in the main paper shows a slice of the high symmetry plane of the background-subtracted diffuse scattering measurement at $300$~mK and zero field. The sample was cooled from a higher temperature to $300$~mK with no interruption within less than an hour (rapid cooling). 
	
	\begin{figure*}
		\centering
		\includegraphics[width=2\columnwidth]{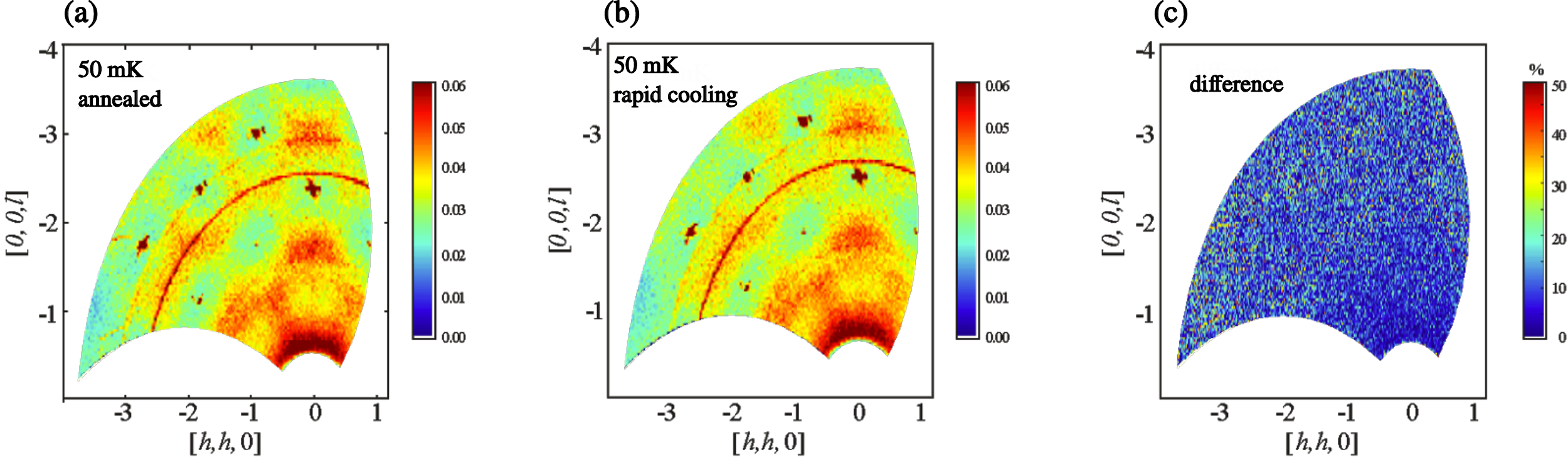}
		\caption{Comparison between the diffuse scattering pattern for the two cooldown procedures: the figure shows the raw data (without any background subtraction) for the S(Q) (both corresponding to cycle 1 of Fig S 1). (a) Diffuse pattern after annealing (point A1 in Fig. S 1). (b) diffuse pattern after a rapid cooling (point R1 of Fig. S 1). (c) subtraction of the two measurements. No significant difference is observed.
		}
		\label{S2}
	\end{figure*}

	We performed neutron scattering experiments fusing fast and slow annealing protocols using the E2 flat cone diffractometer at the BER II research reactor, Helmholtz-Zentrum Berlin. A PG monochromator provided a wavelength $\lambda = 2.39$~\AA, and the diffuse scattering was measured with four $300 \time 300~{\rm mm}^2$ position sensitive detectors. A single crystal sample was aligned with the [hkl]=[100] and [010] directions in the horizontal scattering plane. A dilution insert was used to reach temperatures down to $0.04$~K. We performed two run-cycles,  using the cooling protocols described in Fig.~S\ref{S1} (red: cycle 1; blue: cycle 2). Each cycle consisted of an annealing where the temperature was gradually lowered (cycle 1 for around 8 days, cycle 2 for around 27 days), where the temperature was gradually lowered towards a target at low temperature ($50$~mK in cycle 1 and $300$~mK in cycle 2).  Measurements were performed once these temperatures were reached (A1 and A2 in the figure), called ``annealings''.  This was followed by a series of increases towards fixed temperatures, and finally by a rapid cooling towards the target temperature (in less than 1 hour in both cases), after which the ``rapid cooling'' measurements were taken (R1 and R2) in the figure.
	Data were collected then converted into 3D reciprocal space using the instrument’s TVnexus software. A measurement at saturated field was subtracted (with Bragg peaks in the background being replaced by a suitable region of the scattering in Q-space) to reduce background contamination of the diffuse scattering. 
	
	In addition to the neutron scattering data shown in the main article (cycle 2 in E2 and Corelli), we show raw data of cycle 1 in E2 obtained after a rapid cooling and annealling. Fig.~S\ref{S2}(a) shows the rapid cooling (corresponding to the point R1 in Fig.~S\ref{S1}), with the temperature measured at a thermometer attached to the sample going from $0.8$~K down to $50$~mK in approximately $50$~min. The scattering cross-sections after annealing for approximately 8 days down to the same temperature (corresponding to the point A1 in Fig.~S\ref{S1}) is shown in Fig.~S\ref{S2}(b). No difference was found between the annealed and rapid cooled diffuse scattering patterns, as shown in Figs.~S\ref{S2}(c).

	\section{Numerical simulations and analysis}\label{N1}
	
	Monte-Carlo simulations with the standard single spin-flip Metropolis algorithm were used to study a dipolar spin-ice Hamiltonian that includes exchange terms up to third-nearest neighbors:
	\begin{eqnarray}
	\label{eq:H_comp}
	{\cal H}\   \!\!\! &=& \!\!\! \  Da^3 \sum_{i<j}\left[
	\frac{\Vec{S}_i \cdot \Vec{S}_j}{r_{ij}^3} - \frac{3 \left( \Vec{S}_i \cdot \Vec{r}_{ij}\right) \left( \Vec{S}_j \cdot \Vec{r}_{ij}\right)}{r_{ij}^5}  \right]
	\nonumber \\ 
	\!\!\! &+& \!\!\!  \  J_1\sum_{\left<i,j\right>}  \Vec{S}_i \cdot \Vec{S}_j + J_2 \sum_{\left<i,j\right>_{2}}  \Vec{S}_i \cdot \Vec{S}_j 
	\nonumber \\
	\!\!\! &+& \!\!\! J_3\sum_{\left<i,j\right>_3}  \Vec{S}_i \cdot \Vec{S}_j  
	+ J_3'\sum_{\left<i,j\right>_{3'}}  \Vec{S}_i \cdot \Vec{S}_j 
	\, .
	\nonumber \\
	\!\!\! &\  &
	\end{eqnarray}
	The first term is the dipolar interaction between two spins $i$ and $j$ at a distance $r_{ij}$. $D$ is  the  dipolar  coupling  constant and  $a$ is  the spin nearest-neighbour distance. $D = 1.3224$~K was determined from prior work~\cite{yavors2008dy}. The additional terms account for exchange interactions between first, second and two different third nearest-neighbors, with strengths $J_1$, $J_2$, $J_3$ and $J_3\prime$, respectively (see Fig.~1 in the main text for the definition of the different $J_i$ couplings). 
	
	A cubic box of spins (referred to as supercell), of linear size $L$ nuclear unit cells, containing 16 magnetic sites was used for numerical simulations with periodic boundary conditions. $L=4$, corresponding to 1024 spins was found to be adequate to capture the experimentally observed short-range correlations in a range of temperatures from $100$~mK to $1$~K. 
	
	Initial configurations at each temperature were prepared, and then thermodynamic quantities of interest were calculated in a Monte Carlo ‘production’ run. The system was prepared by starting from a random spin configuration and monitoring its evolution at a fixed target temperature. The total energy, $E$, was recorded at each MC step and analyzed after every $1000$~MC steps. The preparation was finished when the absolute value of the energy difference between the start and end of the $1000$~MC steps window fell below the standard deviation over the same window. This usually took about two cycles, namely about $2000$~MC steps in total. Comparison between different variations in the equilibration condition show that systems prepared in this fashion are reliably in thermodynamic equilibrium above $T_{\rm irr}$. 
	After the preparation stage, the Metropolis evolution in the production run was continued over a period between $10^3$ and $10^5$~MC steps during which samples were taken to calculate quantities such as the structure factor $S(Q)$ or the magnetization $M$. Final quantities were calculating by averaging over approximately $10^3$ independent MC simulations. 
	
	\begin{figure}
		\centering
		\includegraphics[width=\columnwidth]{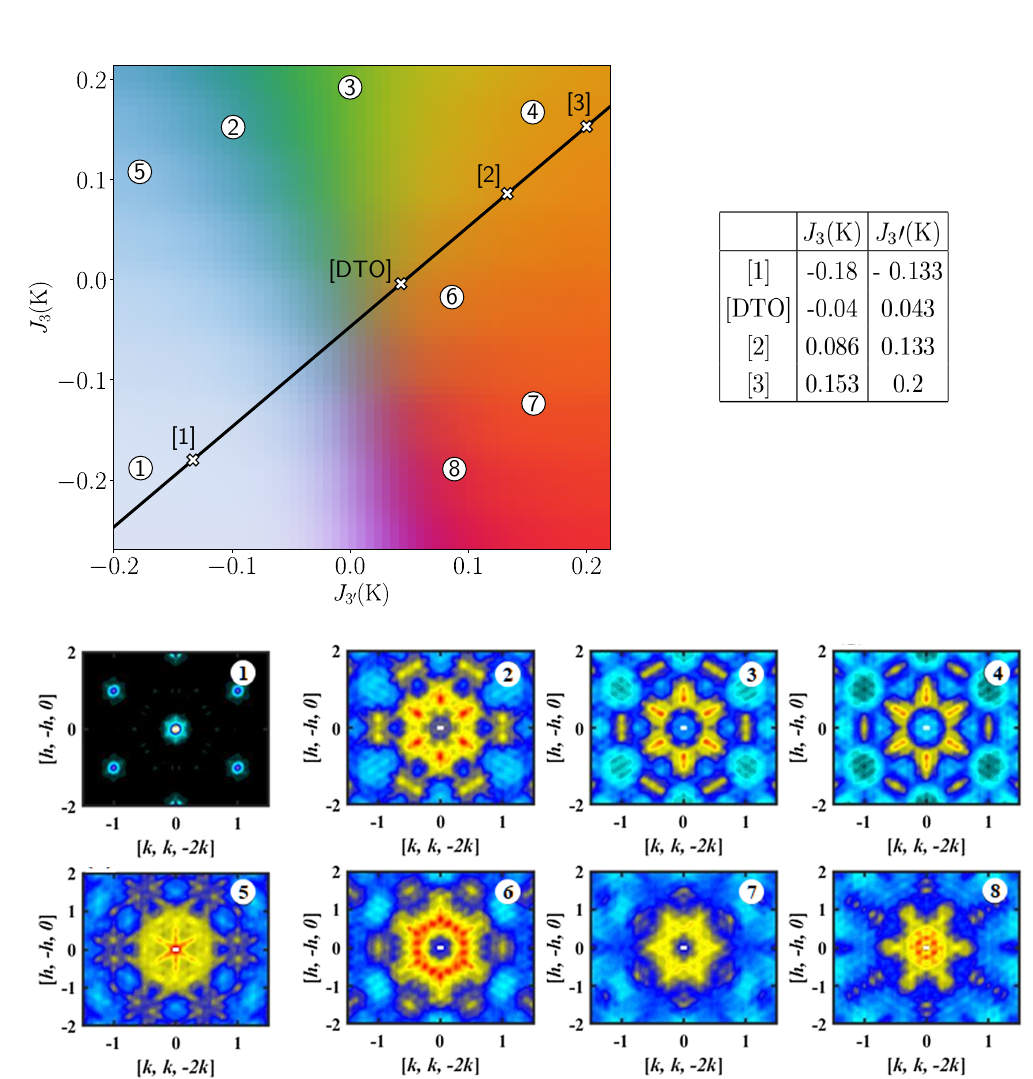}
		\caption{The 2D slice of parameter space at $J_1= 3.41$~K, $J_2 = 0$~K, and $D= 1.3224$~K. The underlying color map shows regions in phase-space characterised by different neutron scattering patterns. Places with the same color have the same $S(Q)$ and the change of color along a specific direction indicates a phase transition/crossover between different structures. The color map was generated by the latent space representation of a non-linear autoencoder (see Ref.~\onlinecite{tennant2021machine} for more details).
			The S(Q) corresponding to selected positions (labeled with numbers) are reproduced here from Ref.~\onlinecite{tennant2021machine} (\copyright IOP Publishing. Reproduced with permission. All rights reserved).     
			The optimal region for DTO is marked by a white cross and the label [DTO]. Three other parameter space points studied in the main article are also indicated with white crosses and the labels [1],[2],[3]. The table gives the exact location in parameter space for these four cases. In the main paper we study the behaviour of the system as we vary the parameters along the straight line $J_3\prime - J_3 = 0.047$~K, marked by the black line.
		}
		\label{S4}
	\end{figure}
	
	Previously, a manifold of solutions simultaneously fitting both $S(Q)$ at $680$~mK and the temperature dependence of the heat capacity was located in parameter space using a machine-learning assisted optimization algorithm by a subset of the current authors~\cite{Samarakoon_2020}. A cross-cut of the ellipsoidal manifold (referred to as the {\em optimal region}) in the plane is shown in Fig.~S\ref{S4} on top of an autonomously generated pseudo phase map. The properties of the Hamiltonian along a straight line in $J_3$  $J_3\prime$ space, given by $J_3\prime - J_3 =  0.047$~K, cutting throught the optimal region was studied numerically. The specific heat as a function of temperature along this line in a fine grid was calculated, see Fig.~3B in the main text. Four parameter sets along this line, including a solution to DTO as listed in Fig.~S\ref{S4}, were used in the numerical investigation of magnetic nanodomain formation, monopole dynamics, and magnetic noise.

	\subsection{Fit to experimental data}
	
\begin{figure}
		\centering
		\includegraphics[width=\columnwidth]{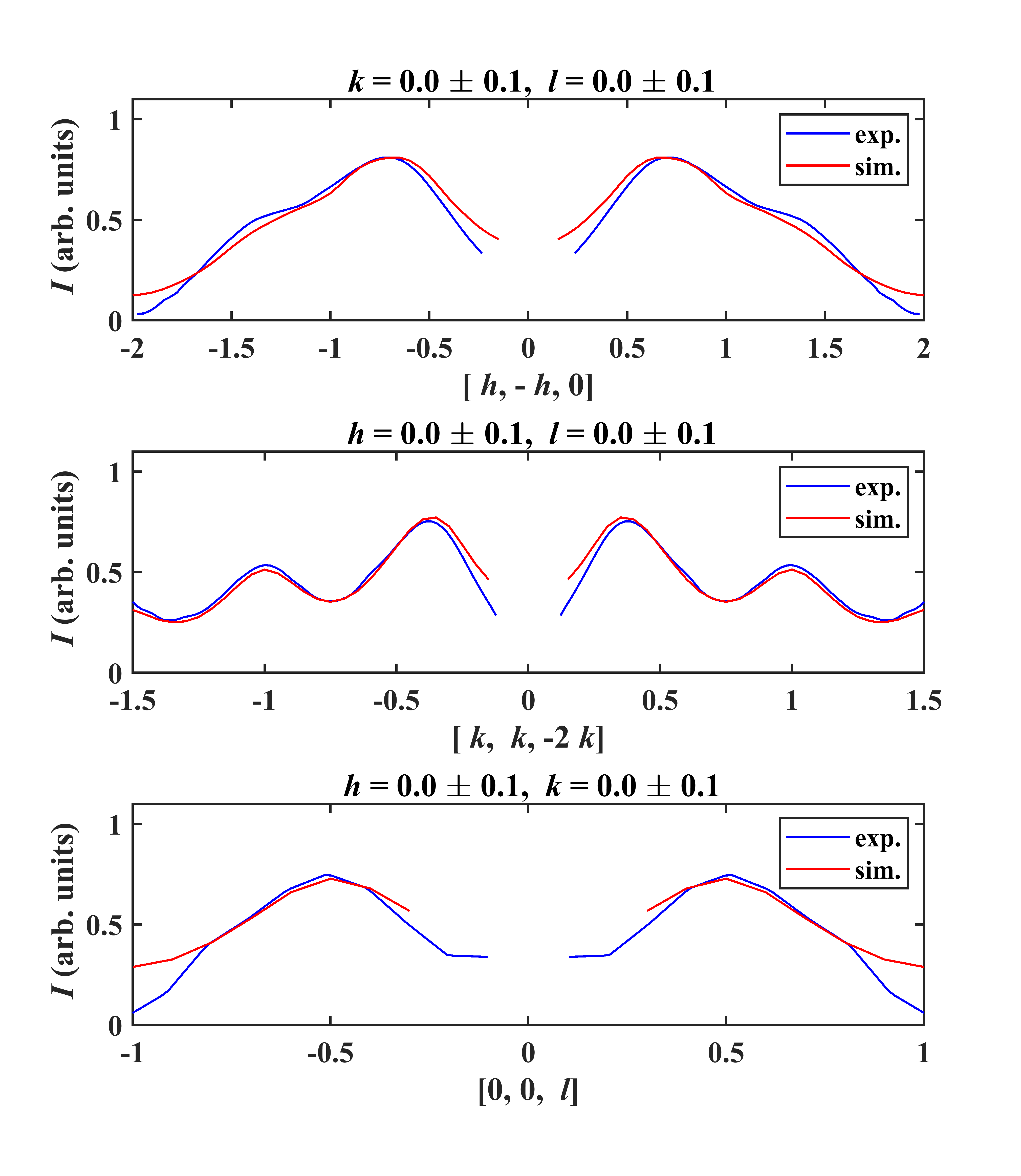}
		\caption{Three line cuts of the same data shown in Fig.~2B in the main text. They are taken along [$h,-h,0$] with $k=l=0$, along [$k,k,-2K$] with $h=l=0$, and along [$0,0,l$] with $h=k=0$. }
		\label{linecuts}
	\end{figure}

	In addition to the comparison shown in Fig 2B in the main text, Figure S.\ref{linecuts} shows three line-cuts to provide an easier quantitative comparison between the simulated neutron diffraction patterns and the experiments.

	\subsection{Visualization of configurations}
	
\begin{figure}
		\centering
		\includegraphics[width=1.1\columnwidth]{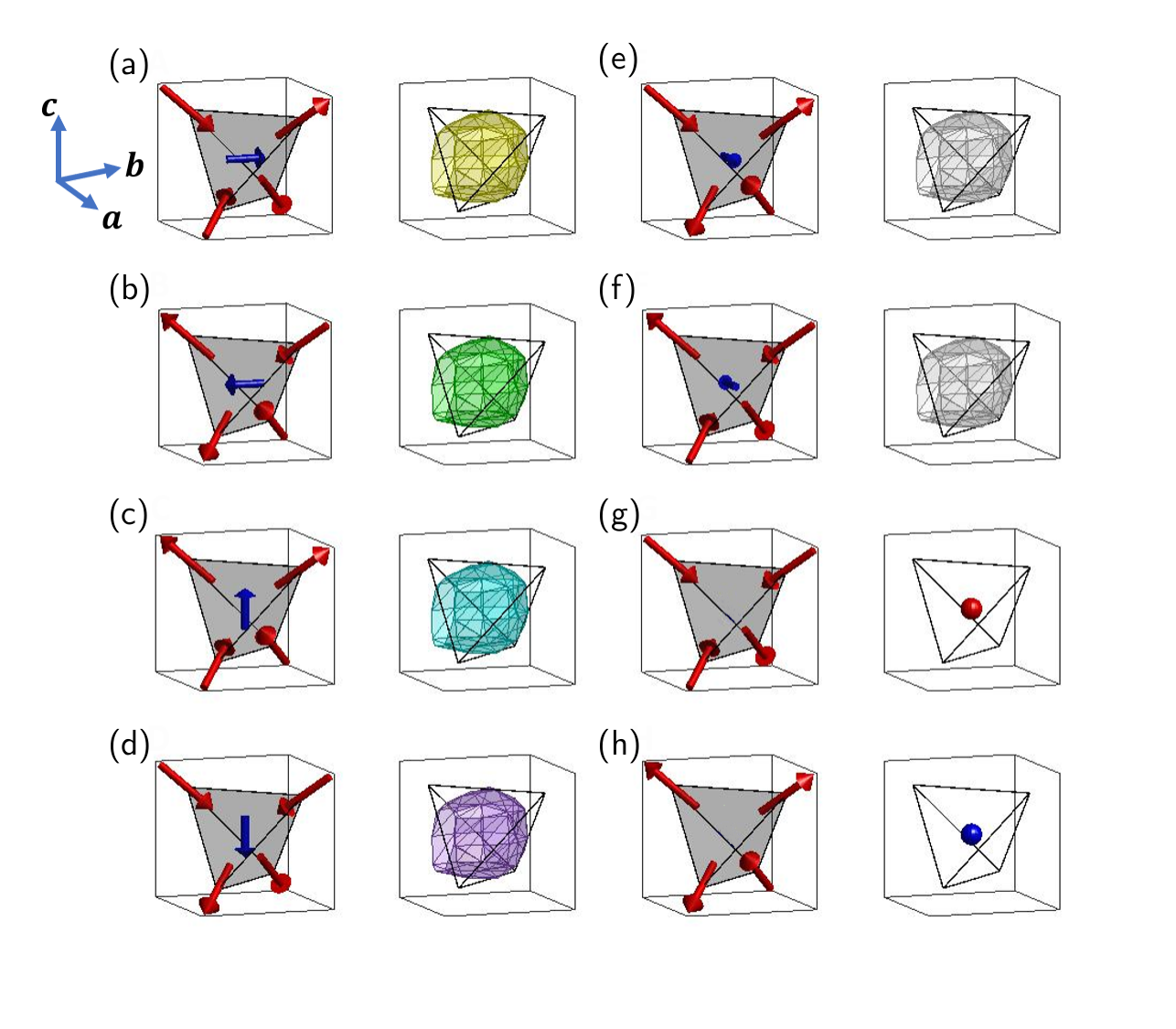}
		\caption{Spin configurations found in the tetrahedra of thermalized spin structures for the considered spin ice Hamiltonian below $1$~K. The spin configurations are eight: six different `2in-2out' configurations with different net moments (A - F), one `3in-1out' (G) and one `1in-3out' (H). The red arrows stand for the Ising spins at the corners of each tetrahedron. The blue arrow at the center shows the net magnetic moment of each motif. The `3in-1out' and `1in-3out' structures are identified as monopoles with opposite signs (there are actually $4$ inequivalent spin configurations for each sign, which are however identified in the monopole charge description). The `2 in - 2 out' motifs have the exact net moment with six different directions along principal crystallographic axes. Hence, we label them in different colors. In a thermalized state, the `2in-2out' motifs have complex superstructures depending on the parameter set. For visualization purposes, in the main text we replace the tetrahedral motifs by the representation shown in the corresponding side panels.
			In this representation, `2 in - 2 out' motifs are shown as surfaces with different colors around the center of tetrahedrons (we only show $4$ out of the $6$ motifs in the main text, hence panels E and F being greyed out). In addition, all `3 in - 1 out' and `1 in - 3 out' structures are shown as red and blue spheres, respectively. }
		\label{S5}
	\end{figure}

	For a better understanding of the spin configurations at low temperatures we choose a depiction in terms of the net moments of the tetrahedra of the pyrochlore lattice. At temperatures below $T<1$~K, only spin-ice configurations (`2in - 2out') and single monopole configurations (`3in - 1out' and `1in - 3out') are found. We replace the tetrahedra with `2in - 2out' configurations by 3D surfaces with six different colors around the tetrahedral centers -- depending on their net moment -- as shown in the side-by panels of Fig.~S\ref{S5}A-F. Spin configurations are replaced by colored volumes at each tetrahedron, constructed such that they smoothly connect, with no overlay, with those of neighboring tetrahedra in the pyrochlore lattice. For simplicity we only show four colors corresponding to four `2in - 2out' configurations. The monopolar configurations ‘3in - 1out’ and ‘1in - 3out’  were replaced by red and blue spheres respectively (see Fig.~S\ref{S5}G and~H). 
	
	\subsection{Videos of simulated monopole dynamics}
	
	We include here a description for each of the four supplementary videos:
	
	Video S1 [File~\texttt{Movie\_S1\_1K.mp4}]: This video contains a visualization of the time evolution of the spin configuration, together with that of  $M$ and $E$ for a fixed temperature of $1$~K. The time evolution of a thermalized spin configuration was calculated for $1.2 \times 10^4$ Monte-Carlo (MC) steps ($\approx 10^7$ single spin-flip attempts) using standard Metropolis dynamics. There were 1024 spin flip attempts between two frames, which is equivalent to the total number of spins in the simulation box. The spin configuration at each MC step was visualised using the convention described above. The pyrochlore lattice is also drawn in dotted lines as a guide. Within the temperature range considered here, a single spin-flip can either move an existing monopole from one center of a tetrahedron to another or create or annihilate a pair of monopoles with opposite sign. The creation of double charges, ‘4in’ or ‘4out’, is too energetically costly at this temperature and it is not observed. Note that both the monopole pathways and the coloured surfaces extend over the three dimensional volume. The perspective of the plot is continuously rotated around the z-axis as time evolves to help the visualisation. The trace of each monopole is shown by lines connecting the centers of the visited tetrahedra, with their color matching that of the originating monopole.

	Video S2 [File~\texttt{Movie\_S2\_680mK.mp4}]: This video contains a visualization of the time evolution of the spin configuration, together with that of $M$ and $E$ for a fixed temperature of $680$~mK. The convention and simulation details are the same as for video S1. The temperature in this case is close to $T_{\rm irr}$. Note that the monopole density and the creation events are very low at this temperature compared to $1$~K. Yet, the monopoles can still equilibrate a significantly large portion of the sample. 
	
	Video S3 [File~\texttt{Movie\_S3\_50mK.mp4}]: This video contains a visualization of the time evolution of the spin configuration, together with that of $M$ and $E$ for a fixed temperature of $50$~mK.The convention and simulation details are the same as for video S1. At this temperature ($T< T_{\rm VGT}$), the monopoles are heavily trapped within the energy barriers introduced by their local environment and can ony hop to neighboring sites. The number of available spin configurations is extremely reduced, and they can no longer equilibrate. 
	
	Video S4 [File~\texttt{Movie\_S4\_Comparison.mp4]}: Comparison of spin  dynamics at three different temperature regimes. Left panel, $T< T_{\rm VGT}$ ($T = 50$~mK), middle panel $T_{\rm VGT}<T\leq T_{\rm irr}$ ($T = 680$~mK) and right panel $T > T_{\rm irr}$ ($T = 1$~K). Unlike the previous videos, the coloured surfaces are not included here. The side-by-side comparison clearly shows the differences in the monopole dynamics and the thermalisation rate at different temperatures.

	\subsection{Quantitative analysis on Monopole pathways}
	
	In order to have a better quantitative description of the characteristics of monopole pathways at varying temperatures, we have calculated a series of relevant quantities. In addition to several standard dynamical and geometrical variables -- such as the average monopole lifetime, $\tau_{\rm life}$, the average maximum displacement, $d_{\rm max}$, the fraction of visited sites, $N_s^{\rm Visited}/N_s^{\rm Tot}$, and the average number of back and forth jumps between neighboring sites, $W$ -- we have also calculated the first Betti number, $N_{\rm Betti}$~\cite{hatcher2002}~\footnote{Betti numbers are mathematical tools used to describe the homology of a given simplicial complex, hence they characterize topological spaces based on their dimensional connectivity~\cite{hatcher2002}. The $k^{\rm th}$ order Betti number ($N_{\rm Betti}^k$) represents the number of $k$-dimensional holes in the complex. In particular, $N_{\rm Betti}^0$ stands for the number of connected components, and $N_{\rm Betti}^1$ the number of one-dimensional or ``circular'' holes. The sequence of Betti numbers $\{ N_{\rm Betti}^0,N_{\rm Betti}^1,\ldots,N_{\rm Betti}^k\}$  characterizes an $n$-dimensional complex. For example, it is $\{ 1,0,0 \}$  for a single point, $\{ 1,1,0 \}$ for a circle, $\{ 1,0,1 \}$  for a sphere and $\{ 1,2,1 \}$ for a torus.}
	
	In the context of monopole motion, each monopole traces a path as it moves through the system, which forms a network embedded in the 3D lattice. As shown in Fig.~3(c) and Fig. 4 in the main text, such networks can include loops when a monopole pathway self-intersects. The frequency with which closed loops occur in the motion of a monopole reflects the correlations in the spin configuration and the complexity of the local energy landscape that the monopole experiences. Calculation of the first Betti number ($N_{\rm Betti}^1=N_{\rm Betti}$), which counts the number of such closed loops in a network, allows for a quantitative analysis of this phenomenon. It can be calculated directly using the number of connected components ($C$), vertices ($V$), and edges ($E$), with $N_{\rm Betti}=C+E-V$. However, due to the large number of pathways used in our analysis -- of the order of $10^5$ -- we used instead the javaPlex package for applied topology~\cite{adams2011javaplex}. 
	
	\begin{figure}[h]
		\centering\includegraphics[width=\columnwidth]{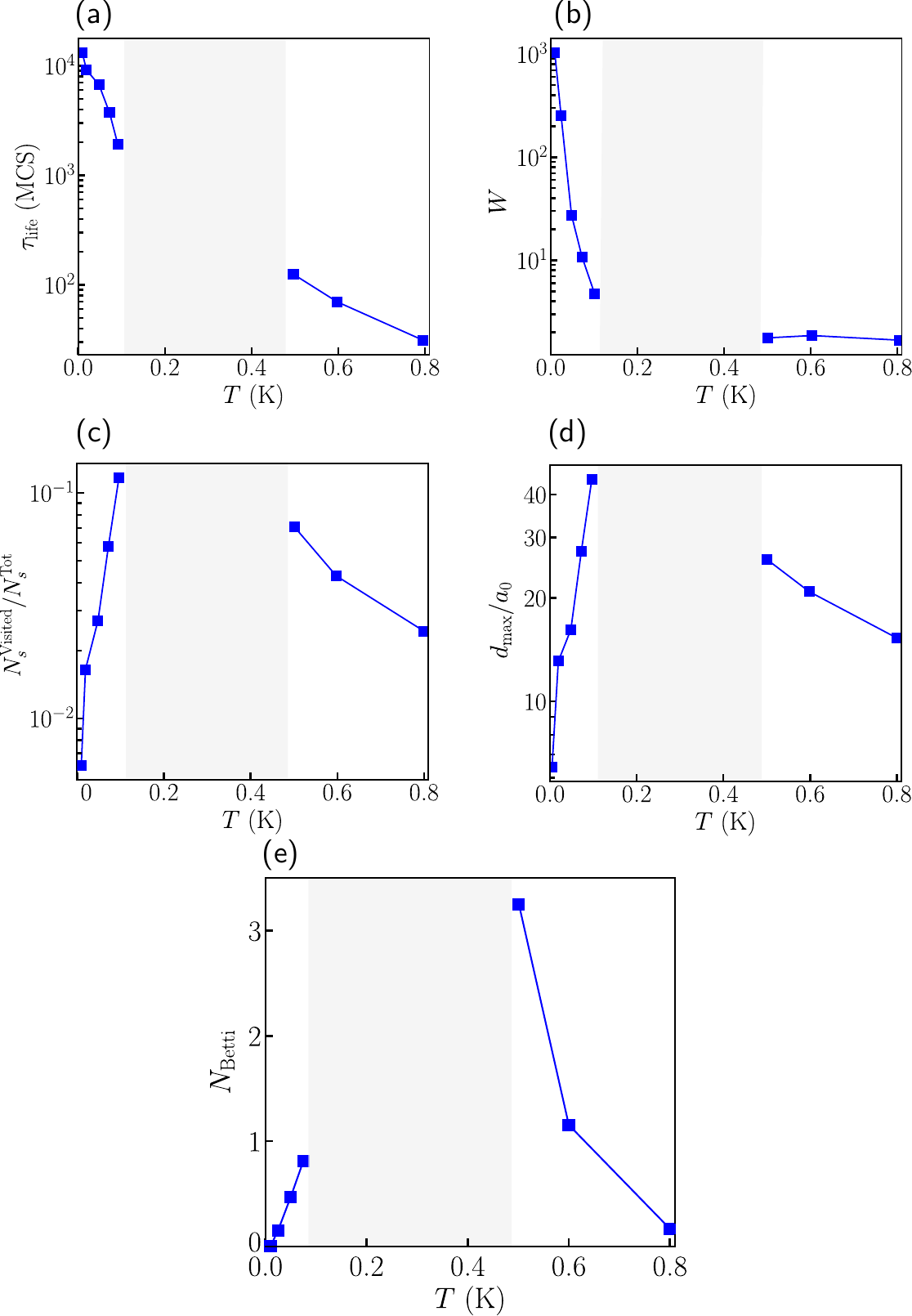}
		\caption{The different mathematical quantities calculated for the monopole pathways as a function of temperature are shown here. (A) averaged monopole lifetime, $\tau_{\rm life}$, (B) average number of back and forth jumps between neighboring sites, $W$, (C) fraction of of visited sites, $N_s^{\rm Visited}/N_s^{\rm Tot}$, (D) maximum displacement from origin, $d_{\rm max}$, measured in terms of the lattice parameter $a_0$, and (E) first Betti number, $N_{\rm Betti}$. The gray area indicates the temperature region where the pathway analysis is not practicable in finite size samples (see text).
		}
		\label{S6}
	\end{figure}
	
	The first step of the analysis required us to identify monopolar excitations and to trace their evolution over time. As described in Sec.~\ref{N1}, for each temperature a finite-size numerical sample is prepared by creating a random initial configuration that is subsequently thermalized. With our chosen procedure (see Sec.~\ref{N1}), we find that above $T_{\rm irr}$ the samples are in thermal equilibrium, excitations abound and there is no difficulty in collecting pathways. 
	Just below $T_{\rm irr}$, monopoles are scarce and it becomes very difficult in a finite-size system ($L=4$) to acquire enough statistics for any meaningful analysis. 
	However, we find that for much lower target temperatures ($T<0.15$K), our preparation leaves behind an appreciable density of quenched monopoles, and we are again in the position to collect enough pathways to produce a statistically relevant analysis. 
	This is why we have data at both low and high $T$, but not in an intervening temperature window (grayed out in Fig.~\ref{S6}). 
	
	The different panels in Fig.~S\ref{S6} show quantitative indicators of the behavior discussed in the main text. Panel A shows the temperature dependence of the average lifetime of a monopole, $\tau_{\rm life}$, measured in MC steps on a logarithmic scale. It increases steadily as the temperature is lowered, a consequence of the drastic reduction in the mobility that makes monopole annihilation events (of monopoles of opposite charge) increasingly rare. This progressive confinement of monopoles as $T$ is lowered is also quantified by $W$, the averaged number of back and forth hops between neighboring sites (panel B), which shows a pronounced raise with decreasing temperature. Two more geometrical indications of the progressive disconnection between monopole pathways and their eventual localization that prevents thermalisation are the fraction of visited sites, $N_s^{\rm Visited}/N_s^{\rm Tot}$, (panel C), and the maximum average distance travelled by a monopole, $d_{\rm max}$, (panel D). Both of these show a maximum at intermediate temperatures, reflecting the competition between the decrease in mobility with temperature, and the decrease in likelihood of annihilation processes due to the decrease in monopole population.
	
	As discussed before, the first Betti number $N_{\rm Betti}$ quantifies the complexity of the monopole pathways in terms of its tendency to generate loops. As the temperature is lowered from $0.8$~K, $N_{\rm Betti}$ raises, (panel E). The progressively longer lived excitations generate complex pathways reflecting the energy landscape they experience. Below $T_{\rm irr}$ the monopoles become increasingly confined, their pathways tend to become short, disconnected segments, and consequently $N_{\rm Betti}$ tends towards $0$. 
	
	
	\section{Relaxation time}
	
	
	%
	%
	
	\subsection{Vogel-Fulcher behaviour in Monte Carlo simulations} 
	
	To complement the experimental study of magnetic relaxation time scales, we performed a similar one on the theoretical models. 
	The relaxation time scale $\tau$ in Monte Carlo simulations was obtained by computing the time dependence of the total magnetisation component along a given crystallographic axis. This is predicated on the identification of the number of Monte Carlo single spin flip attempts with `real time', which has become customary in the field~\cite{Jaubert2009}. 
	We then computed the power spectral density and we fitted it to a Cole-Cole form to extract $\tau$. (We note that the latter is largely insensitive to the anomalous behaviour at higher frequencies and therefore a Cole-Cole choice proved adequate and uncontroversial; fits were performed up to a cutoff frequency of $\omega = 10$ inverse MC steps.) 
	
	We study the full spin ice model, ${\cal H}$, Eq.~\eqref{eq:H_comp}, and two particular cases: the dipolar model, ${\cal H}_{\rm dip}$, when $J_2=J_3=J_3\prime =0$ and the nearest neighbor model, ${\cal H}_{\rm nn}$, when $J_2=J_3=J_3\prime =D=0$. For these we plotted $\tau$ vs. temperature, $T$, over a range of temperatures where thermal equilibrium could be confidently established. We then fitted the data using the forms~\cite{Cavagna2003}: 
	\begin{eqnarray}
	\tau(T) &=& \frac{1}{2} \exp\left(\frac{A}{T}\right) 
	\qquad {\rm Arrhenius} 
	\\
	\tau(T) &=& \frac{1}{2} \exp\left[ \frac{A}{T - T_{\rm VF}} \right] 
	\qquad {\rm Vogel-Fulcher} 
	\, , 
	\end{eqnarray}
	where the prefactor $\tau(T \to \infty)=1/2$ was set a priori, as appropriate for Monte Carlo simulations. The fit was performed using the logarithmic values of $\tau$ and $T$, to avoid giving excessive weight to low-$T$, where the values of $\tau$ are exponentially large. 
	
\begin{figure}[h]
		\centering
		\includegraphics[width=\columnwidth]{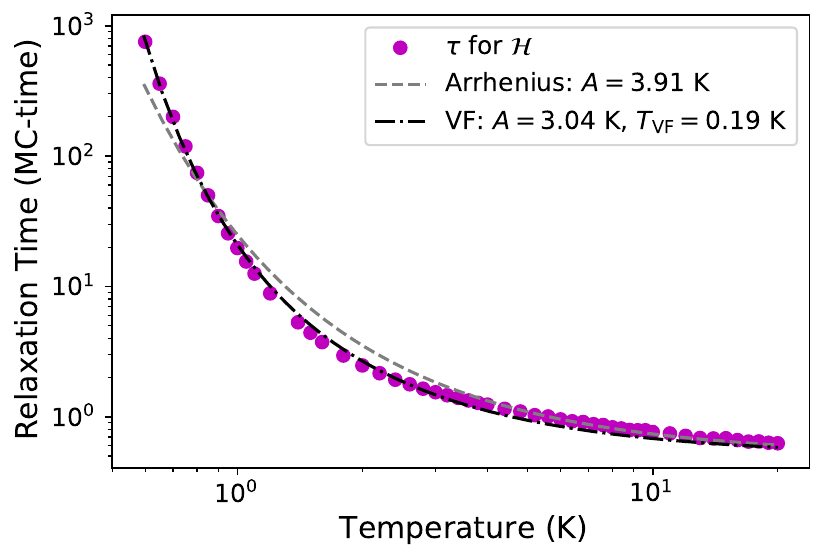}
		\\
		\includegraphics[width=\columnwidth]{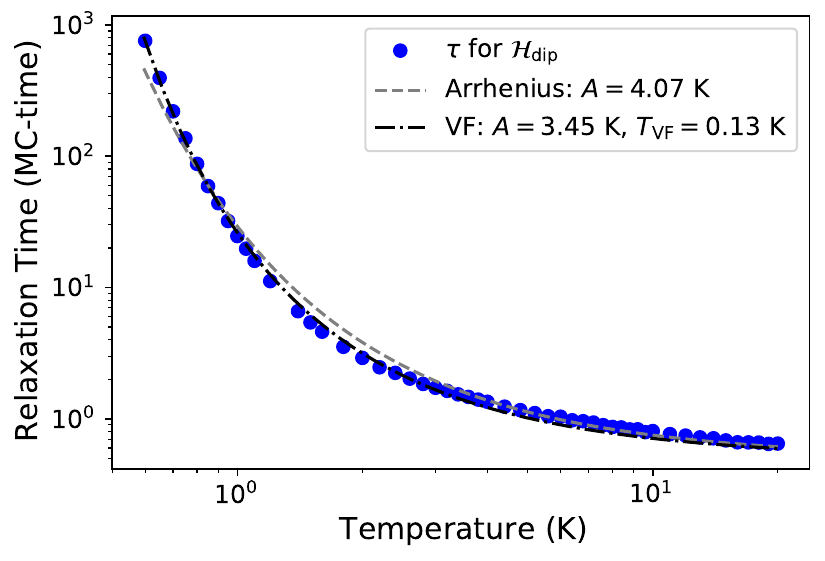}
		\\
		\includegraphics[width=\columnwidth]{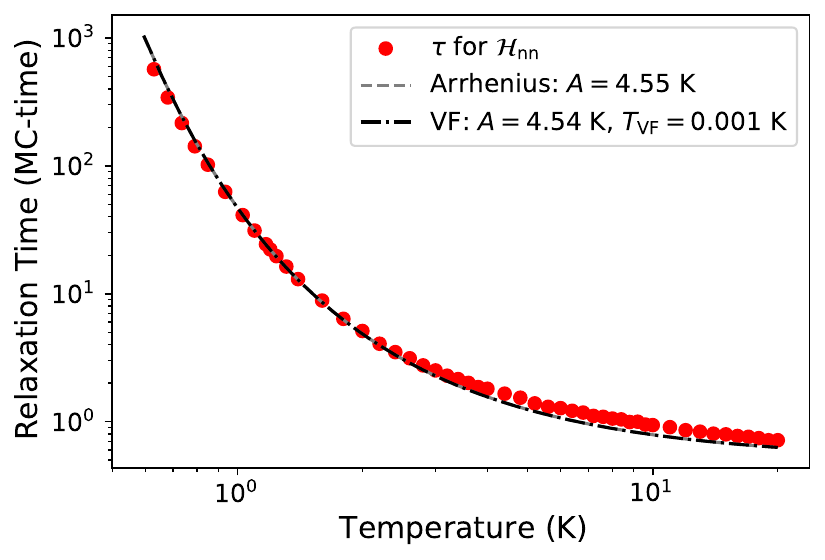}
		\caption{\label{fig:vft}
			Dependence of the relaxation time $\tau$ as a function of temperature $T$, from Monte Carlo simulations in thermodynamic equilibrium for the models ${\cal H}$, ${\cal H}_{\rm dip}$ and ${\cal H}_{\rm nn}$. The dashed and dotted lines show Arrhenius and VF fits over the full temperature range, with fitting parameters given in the respective legends. 
		}
	\end{figure}

	Our results are shown in Fig.~\ref{fig:vft}, and they allow us to make a number of important statements. 
	%
	%
	Firstly, for nearest-neighbour spin ice there is no appreciable difference between the two fitting forms and therefore no clear evidence of departure from Arrhenius behaviour. The situation is different for ${\cal H}$ and ${\cal H}_{\rm dip}$; the Arrhenius form does not manage to achieve a satisfactory agreement with the data, when compared to the VF form, and the departure is more pronounced for the ${\cal H}$ -- where indeed corrections to the Coulomb liquid description are stronger. These observations are consistent with the discussion and interpretation of the experimental data in the main text; furthermore, we similarly find VF divergence temperatures that are consistent with the thermodynamic ordering transition temperatures observed in these models using loop Monte Carlo updates.

	In the literature, an alternative `parabolic' functional form is often contrasted to VF behaviour~\cite{Elmatad2009,Biroli2013}: 
	\begin{eqnarray}
	\tau(T) &=& 
	\frac{1}{2} \exp\left( \frac{B}{T} + \frac{C}{T^2} \right) 
	\, . 
	\end{eqnarray}
	In our simulations, VF and parabolic fits are equally good and essentially indistinguishable (and therefore we chose not to show the latter for simplicity). This suggests that our accessible temperature window is insufficient to discriminate between the two forms (notice indeed that the two agree asymptotically at high temperatures if $B=A$ and $C = A T_{\rm VF}$, which we indeed found to be approximately satisfied by our fitted parameters).

	\bibliography{references}

\end{document}